# Super-Resolution of 3D Micro-CT Images Using Generative Adversarial Networks: Enhancing Resolution and Segmentation Accuracy

Evgeny Ugolkov[a], Xupeng He[b], Hyung Kwak[b], Hussein Hoteit[a]

[a] Physical Science and Engineering, King Abdullah University of Science and Technology (KAUST), Saudi Arabia

[b] EXPEC Advanced Research Center, Aramco, Dhahran, Saudi Arabia

**ABSTRACT**

We develop a procedure for substantially improving the quality of segmented 3D micro-Computed Tomography (micro-CT) images of rocks with a Machine Learning (ML) Generative Model. The proposed model enhances the resolution eightfold (8x) and addresses segmentation inaccuracies due to the overlapping X-ray attenuation in micro-CT measurement for different rock minerals and phases. The proposed generative model is a 3D Deep Convolutional Wasserstein Generative Adversarial Network with Gradient Penalty (3D DC WGAN-GP). The algorithm is trained on segmented 3D low-resolution micro-CT images and segmented unpaired complementary 2D high-resolution Laser Scanning Microscope (LSM) images. The algorithm was demonstrated on multiple samples of Berea sandstones. We achieved high-quality super-resolved 3D images with a resolution of 0.4375 µm/voxel and accurate segmentation for constituting minerals and pore space. The described procedure can significantly expand the modern capabilities of digital rock physics.

## 1. Introduction

Digital Rock Physics (DRP) is a powerful tool that allows for non-destructive analyses and measurements of rock samples through advanced imaging techniques and computational simulations. This technology is used in various industries, such as oil and gas, geothermal energy, carbon capture and storage, mining, and geotechnical engineering. Measuring petrophysical properties directly from digital images offers significant advantages in speed, cost, and the breadth of analyses that can be performed compared to traditional laboratory methods (Saxena et al., 2019). DRP leverages digital representations of rock microstructures, often obtained from micro-CT scans, to assess various petrophysical properties and processes, such as porosity (Pramana et al., 2022; Schepp et al., 2020; Sharma et al., 2021), single- and multi-phase flow simulations with corresponding permeabilities (Alpak et al., 2018; Dvorkin et al., 2003; Fager et al., 2023), phase-field modeling and capillary pressure (Alpak et al., 2019; Cao et al., 2020), nuclear magnetic resonance relaxometry (Arns et al., 2011; Li et al., 2022; Tan et al., 2014), elastic and geomechanical properties (Chalon et al., 2004; Hao et al., 2021; Saad et al., 2019), electrical and thermal conductivity (Shang et al., 2023; Shan-Qi et al., 2012; Yang et al., 2019), reactive transport modeling (CI Steefel and SB Yabusaki, 2000; MacQuarrie and Mayer, 2005; Sadhukhan et al., 2012), and fracture modeling (Wu, 2021).

Despite its broad range of applications, DRP faces several challenges. One of the most critical challenges lies in ensuring that the digital twin, which is a computational model of the rock, is both reliable and representative of the actual physical sample. The accuracy of DRP assessments depends on the image resolution and the segmentation quality across different phases and constituent minerals. However, two inherent trade-offs exist in current micro-CT scanning technologies. The first one is between image resolution and Field Of View (FOV): higher resolution images can capture fine details of the rock's microstructure but often at the expense of a smaller FOV, which may not capture larger scale heterogeneities that could be crucial for representative evaluation (Niu et al., 2020). The second trade-off is between the image contrast and X-ray beam penetration: lower energy X-rays are favored for phase contrast but at the cost of lower penetration depth, making it challenging to image thick or dense samples. Low contrast, in turn, leads to difficulties in accurately identifying the boundaries between solid-fluid and solid-solid phases, such as between the pore space and rock grains or between different rock minerals (Golab et al., 2013). Consequently, current micro-CT measurement practices often yield either sufficiently large FOV but with low-resolution/low-contrast or high-contrast/high-resolution images but with an inadequate FOV.

Recent studies have proposed the use of Machine Learning super-resolution to overcome the problem of insufficient resolution for micro-CT images. To address this problem, two approaches were proposed: either increase the resolution for low-resolution 3D images or reconstruct high-resolution 3D images from a small number of high-resolution 2D images. Wang et al. 2020 applied an Enhanced Deep Super-Resolution Generative Adversarial Network (EDSRGAN) (Ledig et al., 2017; Lim et al., 2017) to achieve 4x increase in resolution for paired low-resolution and high-resolution micro-CT images. Niu et al. (2020) proposed an alternative approach for the same problem by using unpaired images. This approach was enabled by the implemented Cycle-In-Cycle Generative Adversarial Network (CinCGAN) architecture (Zhu et al., 2017). An improved algorithm extended to 3D volumes was proposed by Niu et al., (2022), implementing 3D EDSR architecture with paired low-resolution 3D volume as an input and high-resolution 3D volume as an output. This approach produced similar accuracy as the CinCGAN network with unpaired 2D low-resolution and high-resolution images. Another approach by Anderson et al. (2020) transferred the Transmission X-ray Microscopy (TXM) images to the domain of the Scanning Electron Microscope (SEM) images measured with the same spatial resolution. In a later work (Anderson et al., 2021), the authors used a technique based on TXM FOV and SEM contrast to generate the 3D volumes. Liu and Mukerji (2022) utilized two networks: StyleGAN2ADA (Karras et al., 2020) and CinCGAN. The former network was implemented to augment the small training dataset of high-resolution SEM images, while the latter network was successfully applied to integrate unpaired 2D multiresolution digital rock micro-CT and SEM images. The Super-Resolution 3D volume was generated by Super-Resolving slices in the xy-plane and interpolating along the z-axis. As a result, the scientists achieved images with micro-CT FOV and SEM resolution with a 16x scale factor. Coiffier et al. (2020) introduced the Dimension Augmented GAN (DiAGAN) network, which generated 3D fields from 2D examples. The method introduced a random cut sampling step between the 3D Generator and 2D Discriminator, which allowed the transfer of information about spatial patterns of 2D to the 3D domain. Another work related to generating 3D volumes from a single 2D slice was presented by (Volkhonskiy et al., 2019). The training process was directed to strip the distribution from all the possible 3D models with the conditional Generative Adversarial Network (cGAN) with the central 2D slice being the condition input treated by the autoencoder. In this approach, the trained Generator was conditioned on the input central 2D slice. You et al. (2021) proposed another method to reconstruct 3D digital rocks from 2D cross-section images taken at large constant intervals along the axial direction of the rock sample with the pre-trained Progressive Growing Generative Adversarial Network (PG-GAN). PG-GAN was first trained to generate high-quality gray-scale cross-section images and then to reconstruct the 3D digital rock images by linearly interpolating the inverted latent vectors corresponding to the sparsely scanned images.

Every approach listed above has at least one significant limitation. One major drawback is the limited applicability to 2D images (Liu and Mukerji, 2022; Niu et al., 2020; Wang et al., 2020; You et al., 2021), which makes it impractical to generate reliable 3D volumes without artifacts-bringing interpolation. Another challenge is the reliance on paired and synthetically generated low-resolution and real high-resolution images for the training dataset (Niu et al., 2022; Wang et al., 2020). This training technique is difficult to apply for actual measurements using equipment operating in different scales, for example, low-resolution 3D micro-CT and high-resolution 2D SEM. An additional hurdle involves a challenging alignment between images from different measurement sources, for instance, TXM and FIB SEM (Anderson et al., 2021, 2020). A subsequent barrier is a generation of random 3D volumes without the ability to increase the resolution for a particular pre-defined input low-resolution 3D volume (Coiffier et al., 2020; Volkhonskiy et al., 2019). Furthermore, all the aforementioned approaches are limited to either binary or non-segmented images, which imposes additional constraints for DRP applications.

In this work, our proposed method, inspired by Dahari et al. (2021), employs a 3D Deep Convolutional Wasserstein Generative Adversarial Network with Gradient Penalty (3D DC WGAN-GP) to achieve an 8x super-resolution enhancement for low-resolution 3D micro-CT images. By leveraging segmented 3D micro-CT images and unpaired high-resolution 2D Laser Scanning Microscope (LSM) images, the method overcomes key limitations of prior approaches, such as reliance on paired datasets, interpolation artifacts, and inaccurate segmentation of multivalued images. Unlike earlier models, this approach effectively corrects segmentation errors, refines phase boundaries, and introduces sub-micron features, significantly improving the resolution and accuracy of mineral phase differentiation. The model's ability to process large-scale images using sub-volumes and its independence from paired datasets ensure a practical and scalable solution, making it a transformative advancement for Digital Rock Physics applications.

This manuscript is structured as follows: First, we describe the rock analysis of the Berea sandstone samples, including mineral identification and segmentation methodology, to establish the foundation for image enhancement. Next, the raw image acquisition and segmentation processes are described, covering both the low-resolution 3D micro-CT images and high-resolution 2D LSM images, along with the challenges faced in their segmentation. Following this, the methodology outlines the development of the proposed 3D DC WGAN-GP algorithm, including the preparation of the training dataset using StyleGAN2ADA, the training process, and the architectures of the Generator and Discriminator models. The results and discussion section demonstrates the effectiveness of the model in enhancing resolution, correcting segmentation inaccuracies, and improving phase identification, supported by quantitative and visual metrics. Finally, the conclusion summarizes the advancements made, highlighting the algorithm's potential to expand the capabilities of Digital Rock Physics by enabling more detailed and accurate simulations.

## 2. Methodology

### 2.1. Rock analysis

For this study, we used Berea sandstone, a commonly employed rock in Digital Rock Physics (DRP) research due to its broad applicability and well-documented mineral composition and petrophysical properties (Andrä et al., 2013). Even though the mineral composition of Berea sandstone is widely known and presented in numerous works (Kareem et al., 2017; Lai et al., 2015; Øren and Bakke, 2003), samples from various formations may exhibit differences in the proportions of compositional elements. To confirm the specific composition of our sample, we performed an X-ray Diffraction (XRD) analysis, which identified the primary minerals as quartz (71.1%), feldspars (15.1%), clays (4.8%), and other minor components such as mica, carbonate, and heavy minerals. **Table 1** summarizes the total scale quantification for each mineral group to illustrate the relative abundance of each phase within the Berea sandstone sample. Based on the mineral composition data, we segmented the images into four main groups: pore space, quartz, feldspar, and clay. Less prominent mineral groups were combined with the quartz category for simplicity. This classification approach was applied to both the Low-Resolution (LR) and High-Resolution (HR) images to ensure consistent mineral characterization across the datasets.

**Table 1:** Mineralogical composition of Berea sandstone determined through X-ray Diffraction (XRD) analysis, listing the primary mineral groups, chemical formulas, mass content percentages, and total mass content per group.

| Group | Mineral | Chemical formula | Mass content per mineral, % | Mass content per group, % |
|---|---|---|---|---|
| Quartz | Quartz | $SiO_2$ | 71.1 | 71.1 |
| Feldspar | Plagioclase | $Ca_{0.68}Na_{0.30}(Al_{1.66}Si_{2.34}O_8)$ | 8.4 | 15.1 |
| | Orthoclase | $(K_{0.92}Na_{0.08})((Si_3Al)O_8)$ | 3.1 | |
| | Albite | $(Na_{0.98}Ca_{0.02})(Al_{1.02}Si_{2.98}O_8)$ | 3.6 | |
| Clay | Kaolinite | $Al_2Si_2O_5(OH)_4$ | 2.2 | 4.8 |
| | Illite | $(Na, K)_{1-x}(Al, Mg, Fe)_2(Si, Al)_4O_{10}(OH)_2$ | 0.5 | |
| | Chlorite | $(Mg, Fe)_3(Si, Al)_4O_{10}(OH)_2 \cdot (Mg, Fe)_3(OH)_6$ | 2.1 | |
| Mica | Muscovite | $(K, Ca, Na)(Al, Mg, Fe)_2(Si, Al)_4O_{10}(OH)_2$ | 2.9 | 7 |
| | Biotite | $K(Mg, Fe)_3(AlSi_3O_{10})(OH)_2$ | 4.1 | |
| Heavy minerals | Hematite | $Fe_2O_3$ | 0.2 | 0.3 |
| | Siderite | $Fe(CO_3)$ | 0.1 | |
| Carbonate minerals | Calcite | $Ca(CO_3)$ | 0.6 | 1.7 |
| | Dolomite | $Ca(Mg, Fe)(CO_3)_2$ | 0.5 | |
| | Anhydrite | $CaSO_4$ | 0.6 | |

### 2.2. Raw micro-CT images

Berea rock plugs with a diameter of 6 mm were prepared to achieve high-contrast imaging near the resolution limit of our Tescan CoreTOM X-ray CT scanner. The resulting raw 3D micro-CT images were acquired with a spatial resolution of 3.5 μm, consistent with the equipment's specifications. We used Acquila software for acquisition and reconstruction. The micro-CT scans were conducted with the following parameters: 80 kV, 10 W, 3.5 μm voxel size, and 20 averages. In this study, the term Low-Resolution (LR) refers to these 3D micro-CT images with a resolution of 3.5 μm per voxel. In contrast, the High-Resolution (HR) images, introduced later, correspond to 2D LSM images with a resolution of 0.4375 μm per pixel.

**Figure 1** shows cross-sections of the raw gray-scale 3D micro-CT volume of the Berea sandstone sample, accompanied by three corresponding 2D slices from the center of the volume in the xy-, xz-, and yz- planes. The gray-scale intensity in the 3D image and the corresponding 2D slices provides a representative visualization of the internal structure of the Berea rock. The variations in gray-scale intensity correspond to different mineral phases and pore spaces. These images reveal the composition of the Berea sandstone, capturing the fine details of its microstructure, such as differences in mineral composition and the distribution of pore spaces. While the raw gray-scale images do not explicitly differentiate between specific minerals, they exhibit complex spatial distributions within the rock. These unprocessed images serve as the main foundation for the study, allowing for the application of advanced segmentation techniques to distinguish between phases like pores, quartz, feldspar, and clay. Furthermore, the subsequent application of super-resolution techniques builds on these segmented images, enhancing the resolution and enabling more accurate phase identification, which is essential for precise petrophysical property analysis and flow simulations.

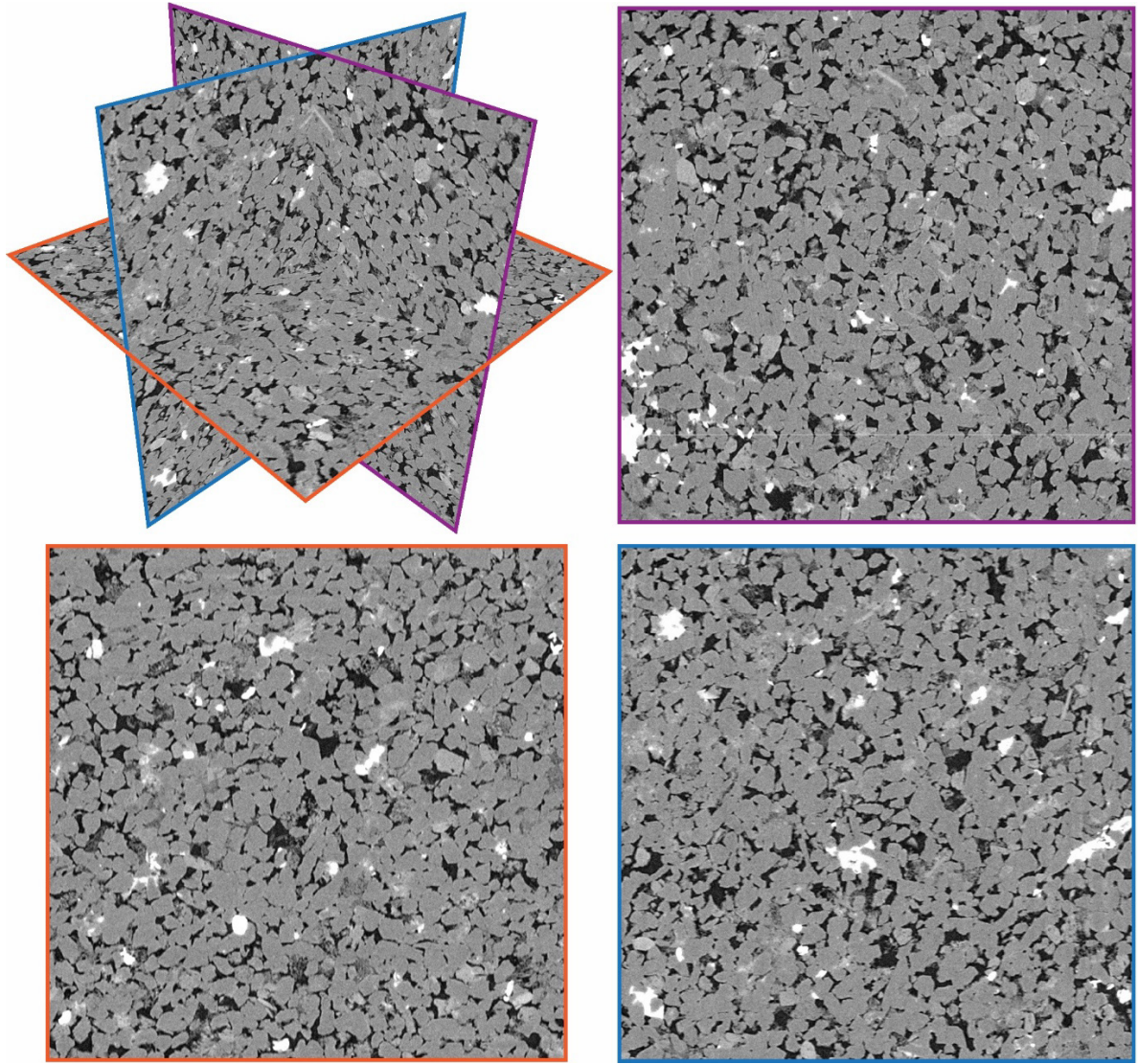

**Figure. 1.** Raw gray-scale 3D micro-CT volume and corresponding 2D slices from the center of the Berea sandstone sample. The 3D volume and its selected 2D slices in the xy-, xz-, and yz-planes display varying intensities in the gray-scale, representing different rock phases within the sample

### 2.3. Segmentation of raw micro-CT images

Segmentation is the process of classifying the gray-scale intensity values in 3D micro-CT images into distinct groups representing different phases of the rock, such as minerals and pore spaces. In this study, the segmentation process was used for identifying and differentiating between key components of the Berea sandstone, including pore space, quartz, feldspar, and clay. This was necessary because raw gray-scale images, while providing a detailed view of the rock's internal structure, do not inherently distinguish between these phases due to overlapping intensity values. For instance, minerals like quartz and feldspar often have similar X-ray attenuation, making it difficult to separate them without segmentation. This segmentation forms the foundation for further image enhancements which subsequently refines phase boundaries and enhances the clarity of sub-micron features.

We segmented the low-resolution 3D micro-CT image into four main groups: pore space, quartz, feldspar, and clay, using Avizo software with the Interactive Threshold functionality (Thermo Fisher Scientific, 2022). The thresholds for the different groups were selected based on visual examination of the minerals' structure and gray-scale intensity. Also, we tested the Watershed Segmentation technique in Avizo, but it didn't provide superior results. **Figure. 2** illustrates the segmented 3D micro-CT volume of the Berea sandstone sample, along with one 2D slice taken from the yz-plane at the center of the volume. In these segmented images, the rock's internal structure is classified into four distinct phases: pore space (black), quartz (dark gray), feldspar (light gray), and clay (white).

Although the segmentation process was overall successful, we faced challenges with accurately segmenting feldspar. This issue was encountered due to the similar electron densities of quartz and feldspar, which led to overlapping X-ray attenuation between these minerals, making it difficult to distinguish them during automatic segmentation (Golab et al., 2013; Tsuchiyama et al., 2000). As a result, the automatic approach was not successful in providing accurate segmentation for the feldspar group: only the brightest spots were highlighted properly, while the rest of the mineral was erroneously attributed to the quartz group (**Figure 2c**). In the following section, we demonstrate how the proposed model can automatically correct these segmentation inaccuracies and refine the identification of feldspar minerals.

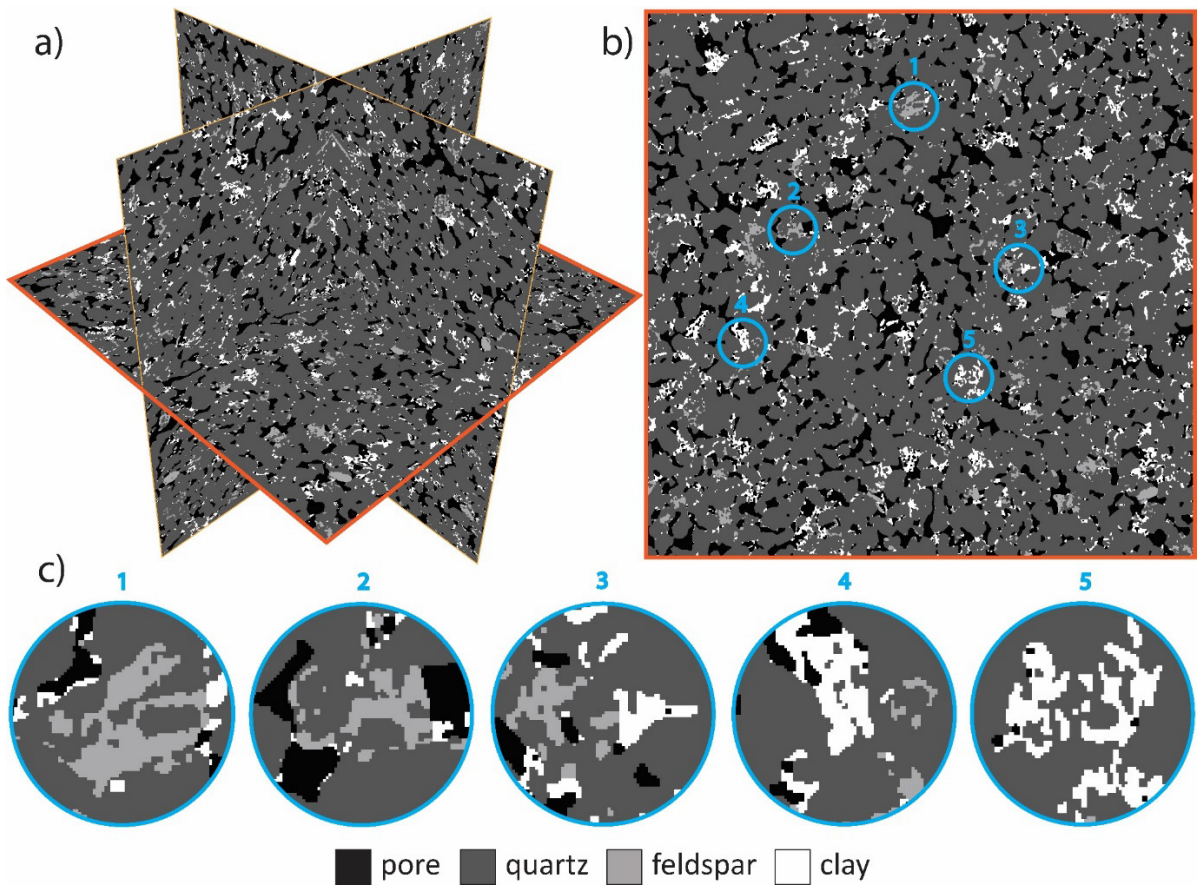

**Figure 2.** Segmentation of the 3D micro-CT volume (a) and a corresponding 2D slice (b) of Berea sandstone. The 3D micro-CT volume and 2D slices (XZ plane) display the segmented structure of the Berea sandstone sample. The segmentation classifies the volume into four distinct groups based on mineral phases: pore space (black), quartz (dark gray), feldspar (light gray), and clay (white). In (b), and (c), we show examples of low-accurate identification of feldspar (1-3) and clay (4-5) using automatic segmentation.

### 2.4. Raw Confocal Laser Scanning Microscope images

Micro-CT images, while effective for capturing the overall structure of a rock sample, often face challenges in limited resolution and distinguishing between minerals with similar X-ray attenuation properties. This leads to segmentation inaccuracies in 3D datasets. High-resolution 2D images using Laser Scanning Microscope (LSM), on the other hand, offer greater spatial resolution, allowing for more accurate mineral identification based on their crystal structure, texture, and composition.

In this study, the High-Resolution (HR) 2D LSM images were acquired with the Zeiss 710 on the thin section stained with Rhodamin B fluorescent ink. This fluorescent treatment offers LSM images a high contrast between the pore and grain states, providing accurate, fast, and automatic pore space segmentation. The measurement parameters were 1 AU, Gain (Master) 415, Digital Offset 1, Power 2.0, wavelength $\lambda$ 514 nm, voxel size 0.4375 µm, and image size of 1948x1948 pixels. **Figure 3a** presents three examples of raw 2D LSM gray-scale images of the Berea sandstone sample. These gray-scale LSM images exhibit varying intensities representing different mineral phases and pore space within the rock.

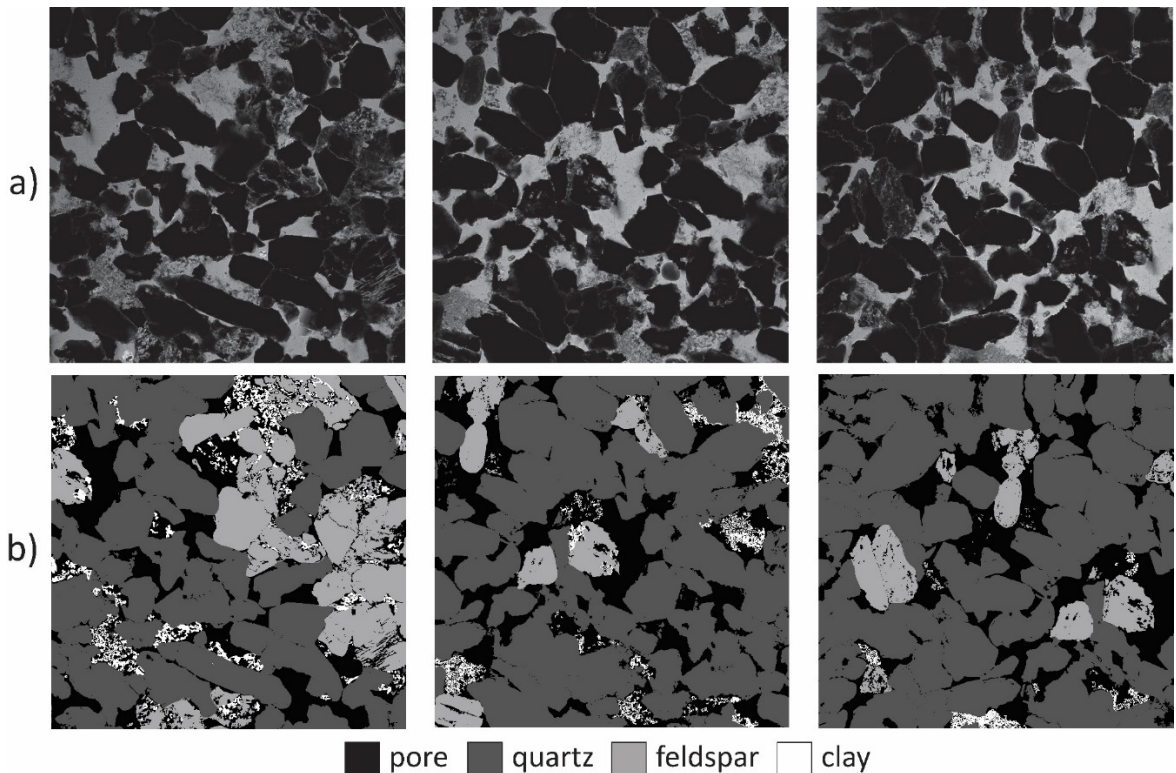

**Figure 3.** Comparison of raw gray-scale and segmented 2D LSM images of Berea sandstone, where (a) shows three examples of raw gray-scale LSM images displaying varying intensities that represent different mineral phases and pore spaces in the rock, and (b) shows the corresponding segmented LSM images, where the rock's composition is classified into four distinct groups: pore space (black), quartz (dark gray), feldspar (light gray), and clay (white). Note that the shown labels correspond to the images in (b) only.

### 2.1. Segmentation of raw Confocal Laser Scanning Microscope images

We segmented the high-resolution 2D LSM images into four main groups, using the Avizo software Segmentation Toolbox. The pores were segmented with the Interactive Threshold method. Similarly to the segmentation issue with the micro-CT images, the minerals exhibited low contrast and were not identified with enough accuracy. Following the recommendations of Golab et al. (2013), we guided our manual mineral segmentation by considering crystal habits, cleavage, and characteristic alteration textures. Feldspar minerals are generally less durable compared to quartz, which has a higher hardness, rating 7 on the Mohs scale, while feldspar typically ranges between 6 and 6.5. This difference in hardness makes quartz more resistant to scratching and weathering, contributing to its greater endurance in various environments. Unlike feldspar, quartz is chemically more stable, making it less susceptible to alteration under chemical weathering processes, which can break down more readily into clay minerals under similar conditions (Alden, 2023). Consequently, the homogeneous minerals were segmented as quartz, while the weathered, fractional, and fractured minerals were segmented as feldspars. The finest fraction in the pore space was segmented as clays. **Figure 3b** displays the same three images after segmentation, where the rock's composition is classified into four distinct groups: pore space (black), quartz (dark gray), feldspar (light gray), and clay (white). This segmentation process enables a more accurate representation of the rock's internal structure by distinctly separating each mineral phase and pore space.

### 2.2. Generation of high-resolution 2D images for training

High-quality, high-resolution datasets are crucial for training machine learning models, especially when tackling challenges related to image resolution enhancement. However, these algorithms typically demand extensive training datasets, often comprising thousands of images, to ensure convergence. Obtaining such a large number of real, high-resolution images is impractical due to the time and labor-intensive nature of manual image acquisition and segmentation. To address this, we utilized a machine learning approach to generate a large number of representative segmented 2D high-resolution images from a limited dataset of 14 ground-truth segmented images. This process allowed us to significantly expand our training dataset, ensuring that the model could learn more accurate representations of the rock's microstructure, particularly for distinguishing between similar mineral phases like quartz, feldspar, and clay. For this purpose, we adapted the StyleGAN2ADA (Style Generative Adversarial Network 2 with Adaptive Discriminator Augmentation) model, introduced by Karras et al. (2020) for generating realistically looking 2D images of people, animals, and other objects with a significantly smaller dataset than commonly used. StyleGAN2ADA is an advanced generative model designed for creating high-quality images from small datasets. It builds on the StyleGAN2 architecture (Karras et al., 2019, 2018), which generates realistic images by learning features in a hierarchical manner. The key enhancement in StyleGAN2ADA is the Adaptive Discriminator Augmentation (ADA), which dynamically applies image augmentations during training to prevent overfitting and improve performance with limited data. This allows StyleGAN2ADA to generate highly realistic images even when the available training dataset is small, making it ideal for applications where collecting large datasets is impractical.

We employed the StyleGAN2ADA algorithm to generate high-resolution segmented 2D images that accurately replicate the properties of real segmented 2D LSM images. For the training process, we initially prepared 14 real segmented images, which were manually checked and corrected to ensure accuracy. To expand the training dataset, we applied image patches of 1024x1024 pixels with a 70% overlap, increasing the dataset size to 349 images. The pixel values in these images were originally in the range of 0 to 3, representing pore (0), quartz (1), feldspar (2), and clay (3). These values were renormalized to fit the StyleGAN2ADA input requirements, translating the 0-3 scale to a 0-255 range: 0 for pore, 85 for quartz, 170 for feldspar, and 255 for clay. This renormalization is essential for the StyleGAN2ADA network, which typically processes images in this 8-bit format.

After training the model, the generated images exhibited pixel values ranging from -30 to 275. To convert these generated values back into four distinct mineral phases, we re-segmented the images by assigning pixel ranges: values between -30 and 45 were mapped to pore (0), 45 to 130 to quartz (1), 130 to 210 to feldspar (2), and 210 to 275 to clay (3). This approach ensured that the generated images closely followed the same mineral phase classifications as the real data.

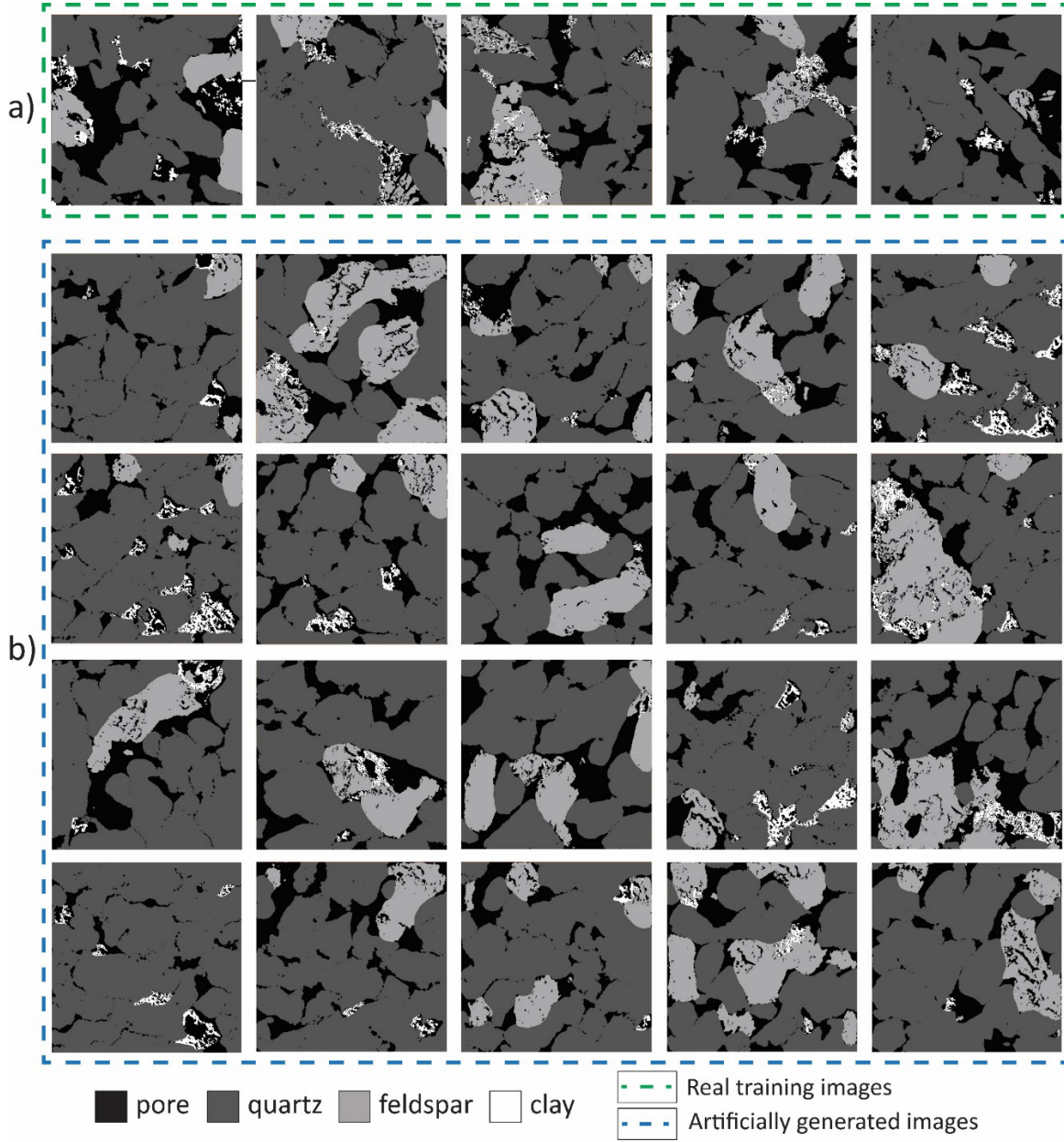

**Figure 4**. Comparison of five actual LSM images (a) and a set of generated segmented 2D LSM (b) images using the StyleGAN2ADA algorithm. The high-resolution generated images provide segmentation of the four distinct

mineral phases: pore space (black), quartz (dark gray), feldspar (light gray), and clay (white), closely replicating the structure and mineral composition of the real images.

A comparison between the real and generated segmented images is shown in **Figure 4**, illustrating the high fidelity of the generated data. Through this process, we successfully produced a dataset of 10000 high-resolution 2D segmented images with a size of 1024x1024 pixels. These generated images were then used to train the 3D micro-CT image processing model, providing a robust and diverse dataset to enhance the machine learning model's performance in segmenting 3D volumes. This expanded dataset was crucial for stable model training and ensuring the algorithm could accurately process and improve the resolution of 3D micro-CT images, effectively correcting segmentation errors and enhancing phase identification.

### 2.3. Proposed super-resolution algorithm for 3D micro-CT images

The proposed algorithm utilizes a Generative Adversarial Network (GAN) to achieve an 8x super-resolution enhancement for the 3D low-resolution micro-CT images. Our adopted GAN algorithm was inspired by the one proposed by Dahari et al. (2021). A GAN is a type of neural network architecture consisting of two competing models: Generator and Discriminator. In the presented case, the 3D Generator, which synthesizes data, competes with the 2D Discriminator, which evaluates authenticity. The Generator aims to create data indistinguishable from real examples, while the Discriminator seeks to distinguish between real and fake generated data. This adversarial process, where both networks are trained simultaneously, encourages the Generator to produce increasingly realistic data.

In the proposed algorithm, the 3D Generator model selects random sub-volume of size $32^3$ voxels from the entire one-hot encoded segmented low-resolution 3D micro-CT training volume of size $1024^3$ voxels. For the training stabilization, the selected sub-volume is concatenated with the Gaussian noise along the channel dimension. The 3D Generator process the selected concatenated sub-volume and generates 3D super-resolution segmented volume of size $256^3$ voxels, increasing the resolution by a factor of 8 from 3.5 μm to 0.4375 μm per voxel. The generated super-resolution volume is automatically corrected for the segmentation errors and incorporates sub-micron features, such as sub-micron pores. Next, the generated 3D super-resolution volume is "sliced" in perpendicular xy-, xz-, and yz-directions into three cross-sections of 2D images, where each cross-section contains 256 images of size $256^2$ pixels.

Thereafter, we train the Discriminator model. In heterogeneous rocks, three separate 2D Discriminators can be trained for each perpendicular plane. As the Berea sandstone being used is homogeneous, a single 2D Discriminator is sufficient for training across all planes. The 2D Discriminator must ensure that each 2D slice extracted from the generated 3D super-resolution volume is statistically similar to the high-resolution 2D LSM segmented images used as ground truth. For each individual stack, the 2D Discriminator model compares the generated slices to real 2D images of the same size, producing the True/Fake Wasserstein (W) loss with Gradient Penalty (GP) (Gulrajani et al., 2017). This loss is used in automatic backpropagation process to update the Discriminator parameters.

Following that, we train the 3D Generator model. The overall structure of the original Low-Resolution image is preserved to avoid unrealistic alterations. To achieve this performance, the generated super-resolution 3D volume is reduced in size 8 times using the nearest neighbor interpolation method, and the Mean Square Error (MSE) loss is calculated between the reduced 3D super-resolution and input 3D low-resolution images. The Wasserstein True/Fake loss for the Generator is calculated and added to the MSE Voxel-wise loss. Subsequently, automatic backpropagation uses this composite loss to update the 3D Generator network. With this approach, the 2D Discriminator refines the 3D Generator's ability to produce realistic 3D super-resolution volumes. Upon completion of backpropagation, the algorithm initiates a new iteration. Graphically, the scheme of the training algorithm for the single iteration is represented in **Figure 5**.

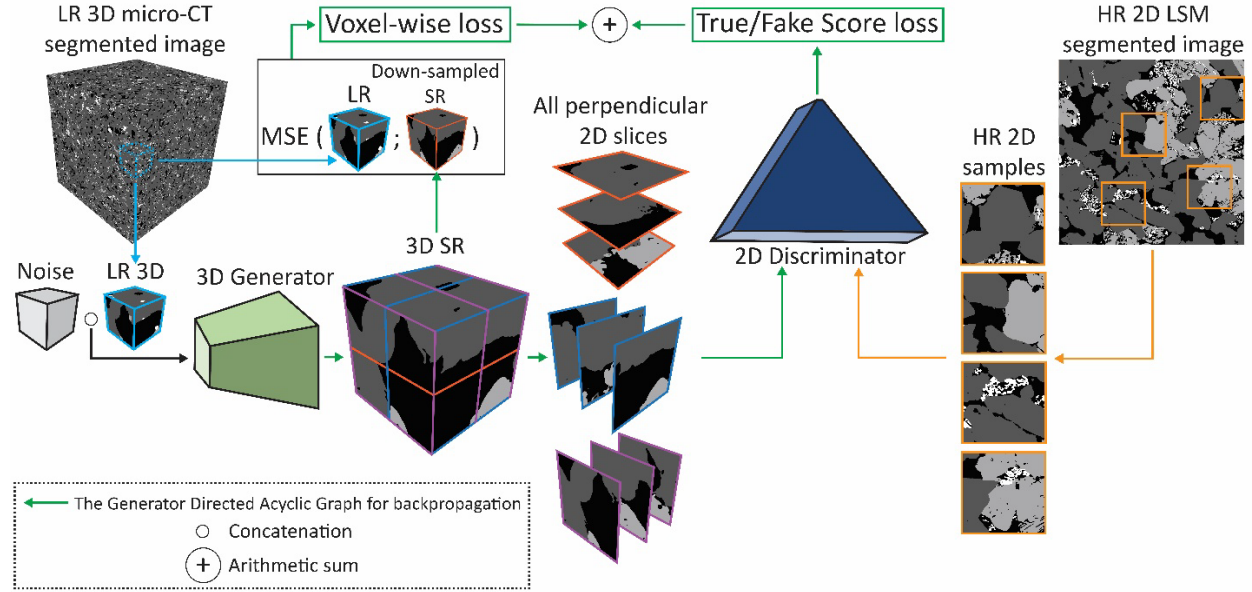

**Figure 5**. Workflow of the 8x Super-Resolution algorithm for 3D micro-CT images. The algorithm begins with the input of a low-resolution (LR) 3D micro-CT segmented image, combined with Gaussian noise to stabilize training. The 3D Generator processes the LR 3D segmented images to produce super-resolution (SR) outputs. The SR image is sliced into 2D cross-sections, which are compared against high-resolution (HR) 2D LSM segmented images using a 2D Discriminator and True/Fake score loss. A voxel-wise Mean Square Error (MSE) loss ensures structural consistency with the original LR image. The combined losses refine the generator through backpropagation, enabling high-resolution and accurate 3D segmentation.

A more detailed description of the algorithm is presented in **Algorithm 1**. The algorithm is not restricted to input volumes of $32^3$ voxels; instead, it can handle arbitrarily large 3D images by processing sub-volumes of the input low-resolution image individually and merging them into a unified output. Such reconstruction process enables the capture of larger scale heterogeneities and facilitates the generation of images with sufficient FOV. The reconstructed 3D images are processed with an iterative 3D median filter to remove the minor artifacts. An important advantage of the presented algorithm is the ability to work with unpaired 3D and 2D images. In other words, the training process involves randomly sampled, non-matching 3D micro-CT images and 2D LSM images. Moreover, since we operate with pre-processed segmented images, the difference in micro-CT and LSM acquisition techniques and image formats doesn't adversely affect the training.

**Algorithm 1:** 8x super-resolution training algorithm for isotropic rocks
Input: $G$ is the 3D Generator function; $D$ is the 2D Discriminator function; $LR$ is the 3D low-resolution training dataset; $HR$ is the high-resolution 2D training dataset; $c$ is the coefficient of the voxel-wise loss multiplication; $\sigma$ is the down sampling function using the nearest neighbor interpolation; $N$ = number of epochs; $n$ = number of iterations per epoch. For clarity and simplicity, the batch operations, optimization processes, and Gradient Penalty (*GP*) parameters have been omitted from the description. Readers are encouraged to consult the project's repository for comprehensive details regarding these parameters, including the specific implementation of the gradient penalty.

Preprocessing:  *HR* ← The data augmentation process: expanding the High-Resolution (*HR*) training dataset by applying all eight possible combinations of mirroring and 90° rotations to *HR*.

    for i = 1, …, *N* do
        for j = 1, …, *n* do
1. *lr* ← uniformly sample a cube of size $32^3$ voxels from *LR*;
2. *z* ← sample a cube of noise of size $32^3$ voxels from normal distribution with mean 0 and standard deviation 1;
3. *lr* ← *concat*(*lr*, *z*) concatenate the low-resolution and noise cubes along the phase dimension;
4. *sr* ← **G**(*lr*) generate a $256^3$ super-resolution volume;
5. $sr_{slices}$ ← *slice*(*sr*) in xy-, yz-, and xz- planes into 3 stacks $sr_{slices\_plane\_k}$. Each stack contains 256 slices of size $256^2$ pixels;

            for k = 1, …, 3 do  *# Discriminator training starts*
6. *hr* ← uniformly sample 128 2D squares of size $256^2$ from *HR*;
7. $l_{gp}$ ← GP($sr_{slices\_plane\_k}$, *hr*, **D**) calculate the Gradient Penalty regularization based on real and fake outputs;
8. $l_D$ ← **D**($sr_{slices\_plane\_k}$) − **D**(*hr*) + $l_{gp}$;
9. Automatic backpropagation and updating the weights of *D* from the loss $l_D$;

           end for *# Discriminator training done*
10. $l_{vw}$ ← *MSE*(*lr*, σ(*sr*)) Voxel-Wise loss between low-resolution and down sampled super-resolution volumes;
11. $l_G$ = 0; *# Generator training starts*

            for k = 1, …, 3 do
12. $l_{G\_plane}$ ← −**D** ($sr_{slices\_plane\_k}$) + $l_{vw} \cdot c$;
13. $l_G = l_G + l_{G\_plane}$;

           end for
14. Automatic backpropagation and updating the weights of **G** from the loss $l_G$;
*# Generator training done*

        end for
    end for
  Output: Trained **G** model.

### 2.4. Model architecture and hyperparameters

The Generator model operates in 3D, where it takes low-resolution 3D segmented micro-CT images with size $32^3$ voxels as input and produces super-resolution 3D segmented images with size $256^3$ voxels. To utilize convolutional layers, all images in the training dataset were one-hot encoded, ensuring that the number of input image channels corresponds to the number of segmentation groups. For training stabilization, the Gaussian noise was concatenated along the channel axis. The Generator model is a sequence of 3D Convolution (Conv3D), Upsample, 3D Transposed Convolution (ConvTr3D), 3D Batch Normalization (BN3D), and ReLU non-linearity layers. The Generator architecture is presented in **Table 2**.

**Table 2:** Architecture of the 3D Generator used in the proposed super-resolution algorithm consisting of several layers, including 3D Convolution (Conv3d), Batch Normalization (BatchNorm3d), Residual Blocks, Upsampling, Transposed Convolution (ConvTranspose3d), and Softmax activation. The model's trainable parameters total 6.77 million, distributed across three GPUs using PyTorch's Pipeline Parallel functionality to optimize memory usage. The table specifies the activation function, output shape, parameter count, and GPU allocation for each layer, highlighting the computational structure and efficiency of the Generator.

| Layer | Activation | Output Shape | Params | GPU |
|---|---|---|---|---|
| Input | - | $5 \times 32 \times 32 \times 32$ | - | |
| Conv3d | - | $512 \times 32 \times 32 \times 32$ | 69.1k | |
| BatchNorm3d | ReLU | $512 \times 32 \times 32 \times 32$ | 1.02k | |
| Residual Block | ReLU | $512 \times 32 \times 32 \times 32$ | 2.36M | GPU 0 |
| Upsample | - | $512 \times 64 \times 64 \times 64$ | - | |
| Conv3d | - | $256 \times 64 \times 64 \times 64$ | 3.54M | |
| BatchNorm3d | ReLU | $256 \times 64 \times 64 \times 64$ | 512 | |
| ConvTranspose3d | - | $128 \times 128 \times 128 \times 128$ | 524.4k | |
| BatchNorm3d | ReLU | $128 \times 128 \times 128 \times 128$ | 256 | GPU 1 |
| Conv3d | - | $64 \times 128 \times 128 \times 128$ | 221.3k | |
| BatchNorm3d | ReLU | $64 \times 128 \times 128 \times 128$ | 128 | |
| Upsample | - | $64 \times 256 \times 256 \times 256$ | - | |
| Conv3d | - | $32 \times 256 \times 256 \times 256$ | 55.3k | |
| BatchNorm3d | ReLU | $32 \times 256 \times 256 \times 256$ | 64 | GPU 2 |
| Conv3d | - | $4 \times 256 \times 256 \times 256$ | 3.5k | |
| Softmax | - | $4 \times 256 \times 256 \times 256$ | - | |

Total trainable parameters **6.77M**

On the other hand, the Discriminator model operates in 2D, which consists of a sequence of 2D Convolution (Conv2D) layers. The Discriminator evaluates 2D slices extracted from the super-resolution 3D outputs by comparing them to high-resolution 2D LSM images. In our configuration, it takes the stacks of high-resolution or super-resolution 2D segmented images with a size of $256^2$ pixels and produces a scalar value. The Discriminator architecture is presented in **Table 3.**

The batch size for the Discriminator network was 128 for high-resolution images and 512 for super-resolution images. The batch size for the Generator network was 2. We also tested batch sizes of 4 and 6 for the Generator model; however, these configurations led to reduced training stability and produced noisier results. The batch size of more than 6 crushed the training with an Out Of Memory (OOM) error. To overcome the problem of high memory demand from the side of Conv3D and ConvTr3D layers, we used the Pytorch Pipeline Parallel (Narayanan et al., 2019) functionality and distributed the layers of the Generator model into three 80 GB A100 GPUs, as shown in **Table 2**. For both the Generator and Discriminator models, we used Adam optimizer with $lr = 10^{-4}$, $\beta_1 = 0.5$, and $\beta_2 = 0.999$. We trained the model with 200 epochs. We used a standard open-source Pytorch library for all parts of this work.

**Table 3:** Architecture of the 2D Discriminator used in the proposed super-resolution algorithm. The model consists of a series of 2D Convolutional (Conv2d) layers with ReLU activations, progressively reducing the spatial dimensions while increasing feature depth. The table provides detailed information on activation functions, output shapes, and parameter counts for each layer.

| Layer  | Activation | Output Shape              | Params |
|--------|------------|---------------------------|--------|
| Input  | -          | $4 \times 256 \times 256$ | -      |
| Conv2d | ReLU       | $16 \times 128 \times 128$| 1.0k   |
| Conv2d | ReLU       | $32 \times 64 \times 64$  | 8.2k   |
| Conv2d | ReLU       | $64 \times 32 \times 32$  | 32.8k  |
| Conv2d | ReLU       | $128 \times 16 \times 16$ | 131.2k |
| Conv2d | ReLU       | $256 \times 8 \times 8$   | 524.8k |
| Conv2d | ReLU       | $512 \times 4 \times 4$   | 2.1M   |
| Conv2d | -          | $1 \times 1 \times 1$     | 4.6k   |

Total trainable parameters **2.8M**

## 3. Results

We demonstrate the model's performance on samples from the Berea sandstone. As input, we took the low-resolution 3D micro-CT segmented image of the Berea sandstone of size $256^3$ voxels and super-resolved it 8 times to the super-resolution 3D segmented image of size $2048^3$ voxels. **Figure 6** shows a comparison between the low-resolution (LR) and super-resolution (SR) segmented 3D micro-CT images of the Berea sandstone. It demonstrates the improvements achieved through the proposed algorithm. The LR segmented image (Figure **6**a) corresponds to the original 3D micro-CT dataset with a voxel size of 3.5 μm. While the segmentation identifies the main phases (pore space, quartz, feldspar, and clay), limitations in resolution lead to overlapping boundaries, misclassifications, and the inability to detect sub-micron features. The SR segmented image (Figure **6**b), generated by the proposed algorithm, enhances the voxel resolution to 0.4375 μm. This process also enables more accurate phase differentiation, better representation of the pore space, and the resolution of sub-micron features that were previously undetectable. Notable refinements include smoother transitions between phases and the correction of segmentation inaccuracies, particularly between feldspar and quartz. Figure **6**c and **6**d show additional cross-sections from the LR and SR models.

**Figure 7** provides a detailed comparison of zoomed-in regions from the LR and SR segmented 3D micro-CT images of the Berea sandstone, shown in Figure **6**. The LR regions (Figure 7a) reveal the limitations inherent to the original dataset, which has a voxel size of 3.5 μm. Coarse phase boundaries are evident, with significant misclassification between minerals such as feldspar and quartz. Additionally, the LR images fail to capture sub-micron features within pore spaces and mineral phases, leading to a lack of detail in areas of high complexity. These limitations hinder accurate representation of the rock microstructure and impact subsequent analyses, such as flow simulations and petrophysical property modeling. In contrast, the SR regions (Figure 7b), generated by the proposed 3D DC WGAN-GP algorithm, achieve a voxel size of 0.4375 μm and demonstrate substantial improvements in both resolution and segmentation accuracy. The boundaries between phases are significantly smoother, and previously undetected sub-micron features, such as intra-grain porosity, are now visible. For instance, Region 1 shows enhanced separation of quartz and feldspar phases, with reduced overlap and misclassification; Region 2 reveals finer pore structures within clay and feldspar regions, introducing sub-micron flow paths critical for accurate flow simulations; Region 3 highlights more realistic and continuous structures in the pore space, correcting coarse and jagged boundaries seen in the LR image.

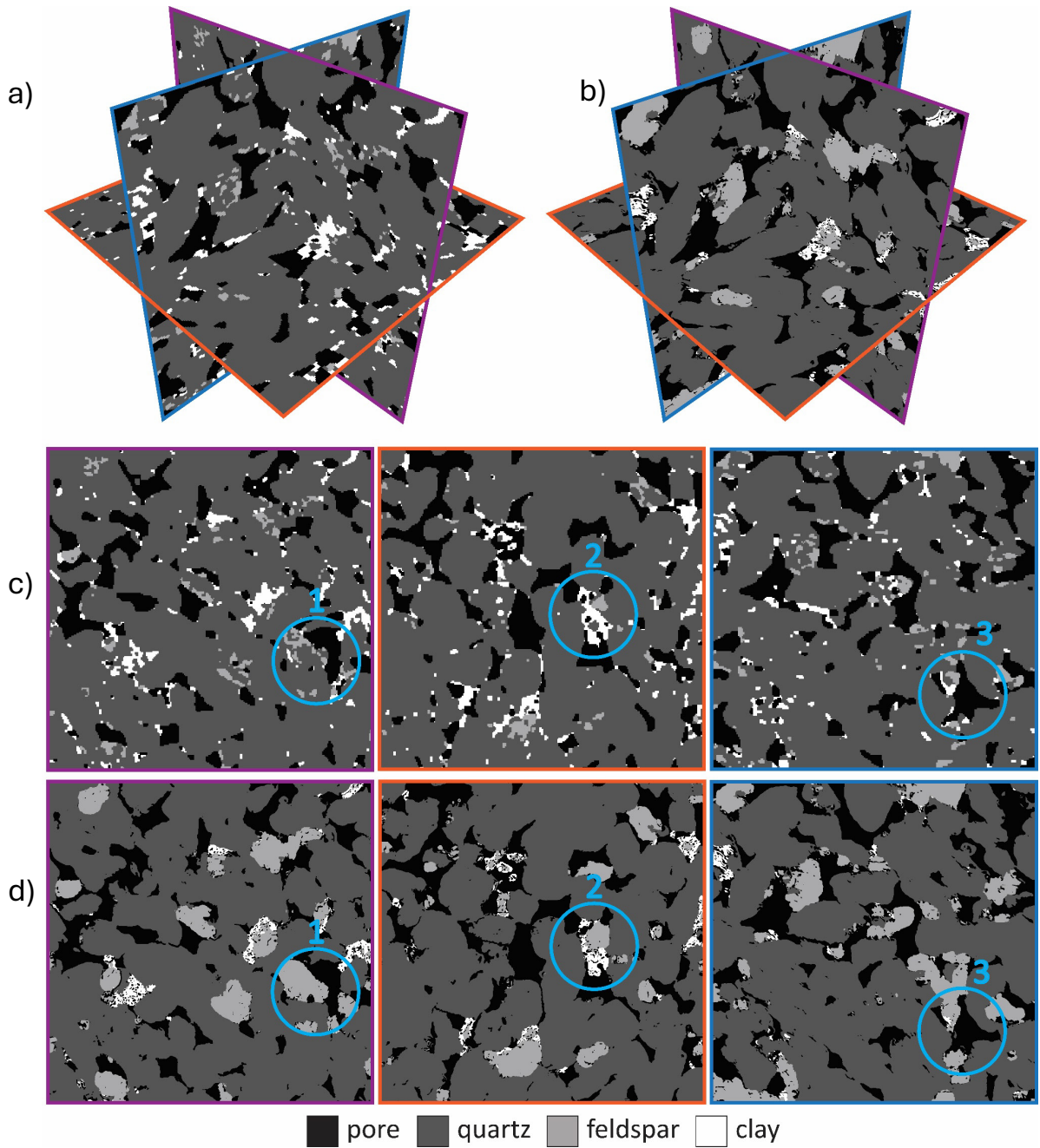

**Figure 6.** Comparison of low-resolution and super-resolution segmented 3D micro-CT images of Berea sandstone. (a) The original Low-Resolution 3D micro-CT segmented volume with a voxel size of 3.5 μm is shown, and (b) Super-resolution (SR) segmented image demonstrating the enhanced 3D segmented volume generated by the proposed algorithm with a voxel size of 0.4375 μm. (c) low-resolution and (d) super-resolution show 2D cross-sections from the 3D models (a) and (b), respectively.

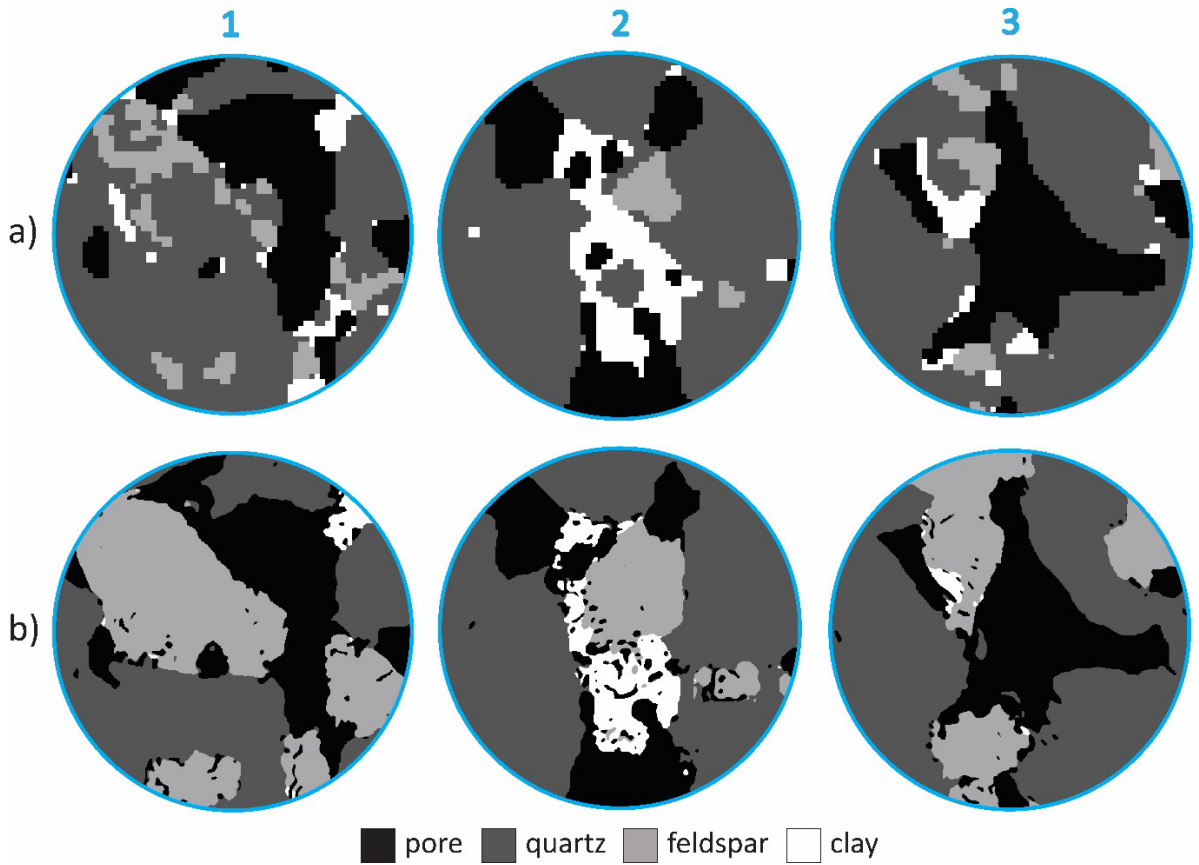

**Figure 7**. Comparison of zoomed-in regions from the low-resolution (LR) and super-resolution (SR) segmented 3D micro-CT images of Berea sandstone, corresponding to areas in Figure 6.

To quantitatively assess the refinements achieved in the 3D super-resolution (SR), and how it compares to the 2D high-resolution (HR), and 3D low-resolution (LR) images, we computed three critical metrics for the pore space: volume fraction, relative surface area, and the two-point correlation function. For all three metrics, 256 patches of size $512^2$, $512^3$, and $64^3$ were randomly sampled from the HR, SR, and LR datasets, respectively. The results validate the effectiveness of the proposed algorithm in improving resolution, segmentation accuracy, and the representation of fine structural details, particularly in challenging areas like pore spaces and phase boundaries. These metrics provide insights into the accuracy of segmentation and the ability of the super-resolution algorithm to resolve fine structural details, as follows:

− *Volume Fraction*: This metric measures the ratio of voxels assigned to a specific group (e.g., pore space, quartz, feldspar, or clay) relative to the total voxels within the region of interest. As shown in **Figure 8a**, the 3D SR results align more closely with the 2D HR images than the LR images, reflecting improved segmentation accuracy and resolution.
− *Relative Surface Area*: This metric quantifies the fraction of interfacial surfaces between two groups (e.g., pore/quartz, pore/clay) relative to the total surface area within the region. As illustrated in **Figure 8b**, the SR images demonstrate smoother and more accurate phase boundaries, with surface area values comparable to those in the HR dataset, particularly for pore-to-quartz and pore-to-clay interfaces.

- *Two-Point Correlation Function*: This statistical measure evaluates the probability that two points, separated by a given distance, belong to the same phase. **Figure 9** depicts the mean and variance of the two-point correlation function for each dataset. The SR results exhibit a closer match to the HR images across all distances, indicating the algorithm's ability to restore spatial correlations and resolve sub-micron features within the pore space.

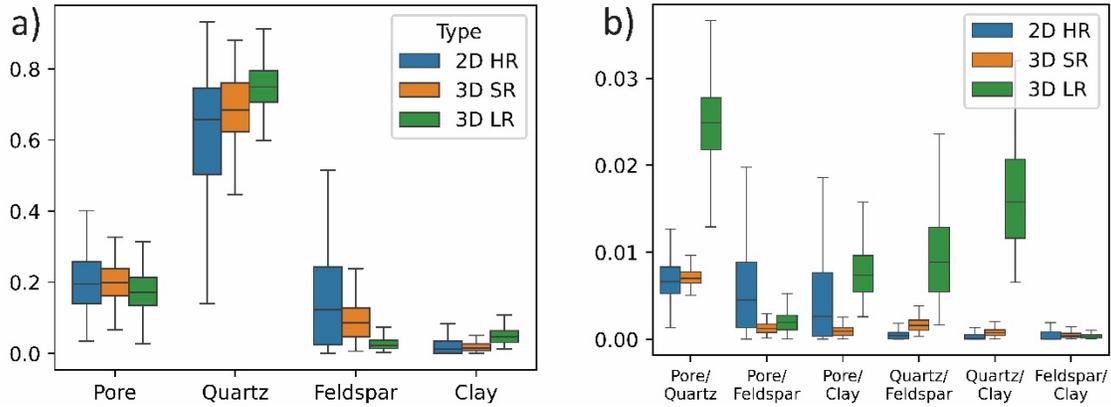

**Figure 8.** Comparison of segmentation metrics for 2D high-resolution (HR), 3D super-resolution (SR), and 3D low-resolution (LR) images. (a) *Volume Fraction*: SR aligns closely with HR, showing improved segmentation over LR. (b) *Relative Surface Area*: SR matches HR with smoother phase boundaries and better pore-phase interfaces.

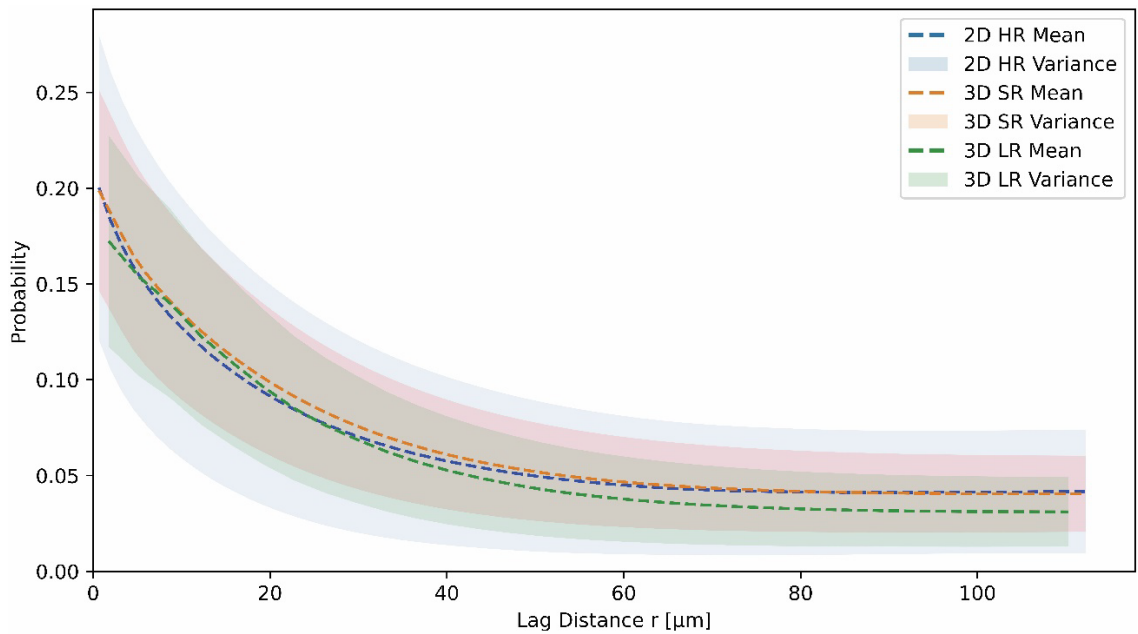

**Figure 9.** Mean-variance plot of the two-point correlation function for 2D high-resolution (HR), 3D super-resolution (SR), and 3D low-resolution (LR) images. The SR results closely match the HR dataset across all lag distances, demonstrating improved spatial correlation and resolution compared to the LR images.

## 4. Discussion

The proposed algorithm significantly enhances the resolution of 3D micro-CT segmented images, increasing the voxel resolution from 3.5 μm to 0.4375 μm, representing an 8x improvement. This advancement addresses critical limitations of low-resolution imaging in Digital Rock Physics, enabling a more accurate representation of rock microstructures.

The term "resolution" in imaging technologies can have varying interpretations, including voxel size, detector pixel size, spatial resolution, sensitivity to density variations, or dimensional analysis accuracy (Rigaku, 2024). In this study, "resolution" specifically refers to voxel size and should not be confused with spatial resolution. Spatial resolution, defined as the minimum resolvable separation between high-contrast objects, depends on voxel size and the Point Spread Function (PSF), which describes system blurriness. Additional factors such as contrast, Signal-to-Noise Ratio (SNR), and artifacts, including the partial volume effect, further influence spatial resolution (Rigaku, 2024).

Using some industry guidelines, a feature with size *L* can be identified if the voxel size within *L/5* to *L/2*, while the shape and size of the feature can be estimated only when the voxel size is within *L/20 ~ L/5* (Rigaku, 2024). Considering our problem, the size *L* of the feature may be defined as an effective radius of any pore or grain. Accepting the least strict parameters from this consideration, we can say that the achieved super-resolution 3D images with voxel size 0.4375 μm can identify the presence of the features of size 0.875 μm while quantifying in detail the features of size up to 2.1875 μm.

The algorithm addresses two key challenges. First, it resolves sub-micron features, such as pores, that are undetectable at low resolution. Quantitative validation of this enhancement is shown in Figure 8a, where the pore volume fraction increases in the super-resolution (SR) images compared to the low-resolution (LR) images. This increase corresponds to a decrease in the clay volume fraction, as sub-micron pores previously misclassified as clay are correctly identified (illustrated in Figure 7 (2)). Additionally, the two-point correlation function (Figure 9) shows a significant correction at longer distances, confirming the improved spatial accuracy of the SR images. Visually, Figure 6 demonstrates how the SR algorithm transforms homogeneous clay and feldspar regions into well-resolved particles with realistic intra-grain porosity, adding critical flow paths that can enhance flow simulation accuracy in digital twins. Furthermore, the refined boundaries between groups, as shown in Figure 8b and visually in Figures 6 and 7 (3), reduce surface roughness inaccuracies and positively impact simulations involving grain surface interactions, such as Nuclear Magnetic Resonance (NMR) Random Walk methods.

Second, the algorithm effectively resolves segmentation inaccuracies, particularly for minerals with similar electron densities, such as feldspar and quartz. As seen in Figures 6 and 7 (1), regions initially misclassified as feldspar are corrected, resulting in more realistic representations that align with high-resolution microscopy images (Figure 4). Figure 8b further illustrates the alignment of relative surface areas between quartz and feldspar phases in SR images with the values derived from the high-resolution dataset. These improvements in mineral differentiation are critical for digital simulations of properties such as elasticity, thermal conductivity, and electrical conductivity.

## 5. Conclusion

In this work, we introduced a novel super-resolution algorithm for segmented 3D micro-CT images, achieving an 8x enhancement in resolution. The voxel size was refined from 3.5 μm to 0.4375 μm, enabling the detection and representation of sub-micron features and correcting segmentation inaccuracies. The algorithm effectively addresses key limitations in current imaging techniques by leveraging unpaired datasets (3D micro-CT and 2D Laser Scanning Microscope (LSM) segmented images), making the acquisition of training data convenient and independent. The high-resolution dataset was manually segmented and significantly expanded using the StyleGAN2ADA functionality, ensuring robust and diverse training data. The model training, performed using the WGAN-GP algorithm with a 3D Generator and 2D Discriminator, demonstrated the ability to generate super-resolved images of arbitrary sizes. The resulting images combine the Field of View (FOV) of micro-CT imaging with the resolution of LSM imaging, significantly advancing the capabilities of Digital Rock Physics (DRP) by providing a more representative and detailed simulation domain. The presented algorithm offers high potential for DRP applications, including more accurate flow simulations, petrophysical property estimation, and elastic property modeling. By refining phase boundaries, incorporating sub-micron details, and improving segmentation accuracy, the algorithm enhances the fidelity and utility of 3D digital twins.

Looking forward, we aim to address several challenges and extend the applicability of this work. First, the use of standard 3D convolutional layers presents memory limitations for larger scale factors, such as 16x, which resulted in Out-Of-Memory (OOM) errors even with Pipeline Parallel implementations. Second, we plan to evaluate the algorithm on a diverse range of rock types, including tight sandstones, where the resolution problem is particularly critical. This will help validate the algorithm's adaptability and performance across varying rock microstructures. Third, we will assess the impact of super-resolution on the accuracy and outcomes of various DRP simulations, including flow modeling, elastic property analysis, and surface roughness calculations.

In summary, the proposed super-resolution algorithm not only resolves the limitations of low-resolution imaging but also lays a strong foundation for future advancements in Digital Rock Physics, enabling more accurate and reliable analyses across a wide range of applications.

**Code and data availability**

This study utilized two codes, both available for download at the specified GitHub repositories.

1. **Code Name**: 8x_Super-Resolution

    - *Contact*: For modified functionality and rock-related applications, contact *evgeny.ugolkov@kaust.edu.sa*. For the original source, contact *a.dahari@imperial.ac.uk*. Refer to the GitHub repository's Readme file for contribution details and credits.
    - *Hardware Requirements*: This work utilized three A100 80GB GPUs in a Pipeline Parallel implementation. For smaller inputs, a single GPU (minimum v100 32GB) may suffice, with adjustments as necessary for specific applications.
    - *Programming Language*: Python.

- *Software Requirements*: Avizo software is recommended for segmentation, but any software producing suitable image formats can be used. The required Python libraries are detailed in the provided environment file.
- *Program Size*: 217 MB, including the training dataset and evaluation volume.
- *Repository*: https://github.com/EvgenyUgolkov/8x_Super-Resolution.

2. **Code Name**: StyleGAN2ADA-for-HR-dataset

- *Contact*: For applications related to rock datasets, contact *evgeny.ugolkov@kaust.edu.sa*.
- *Hardware Requirements*: At least one v100 32GB GPU.
- *Programming Language*: Python.
- *Software Requirements*: Avizo software is suggested for segmentation tasks.
- *Program Size*: 15.9 MB, including the training dataset.
- *Repository:* https://github.com/EvgenyUgolkov/StyleGAN2ADA-for-HR-dataset.

**Acknowledgements**

We express our deep gratitude to Dr. Mohsin Ahmed Shaikh and Dr. Rooh Khurram from the KAUST Supercomputing Core Laboratory (KSL) for their help with the work on the IBEX supercomputer cluster and numerous other problems which are too long to list. We express deep gratitude to Domingo A. Lattanzi Sanchez from the KAUST petroleum engineering core lab for precious advice regarding tuning the micro-CT parameters for high-quality image acquisition. We sincerely thank Jeffery Carpenter, the lab technician at the KAUST petroleum engineering core lab, for his generous assistance in preparing the thin sections. We want to express our gratitude to Dr. Ivan Skorohodov for insightful discussions on the StyleGAN2ADA algorithm and his valuable comments on its application to segmented LSM images.